\documentclass{article}
\usepackage[preprint]{colm2026_conference}

\usepackage{microtype}
\usepackage{booktabs}
\usepackage{graphicx}
\usepackage{subcaption}
\usepackage{multirow}
\usepackage{tabularx}
\usepackage{array}
\usepackage[table]{xcolor}
\usepackage{amsmath}
\usepackage{amssymb}
\usepackage{mathtools}
\usepackage{amsthm}
\usepackage{pifont}
\usepackage{tcolorbox}
\usepackage{bm}
\usepackage{bbm}
\usepackage{wrapfig}
\usepackage{hyperref}
\usepackage{url}

\usepackage{lineno}

\definecolor{darkblue}{rgb}{0, 0, 0.5}
\hypersetup{colorlinks=true, citecolor=darkblue, linkcolor=darkblue, urlcolor=darkblue}

\usepackage{xspace}
\newcommand{\algname}{\textsc{KT4EQG}\xspace}

\newcommand{\aspace}{\hspace{0.5em}}

\title{KT4EQG: Personalized Exercise Question Generation via Knowledge Tracing}

\author{
Xinyi Gao$^1$, \aspace
  Qiucheng Wu$^1$, \aspace
  Lu Ding$^2$, \aspace
  Q.Vera Liao$^{^3}$, \aspace
  Kaizhi Qian$^4$, \aspace
  Ying Xu$^5$, \aspace\\
  \textbf{Shiyu Chang}$^1$, \aspace
  \textbf{Yang Zhang}$^4$ \\[0.5em]
  $^1$University of California, Santa Barbara \\
  $^2$University of South Alabama \\
  $^3$University of Michigan \\
  $^4$MIT-IBM Watson AI Lab \\  
  $^5$Harvard University \\[0.5em]
  \texttt{xinyigao@ucsb.edu}
}

\begin{document}

\ifcolmsubmission
\linenumbers
\fi

\maketitle
 \begin{abstract}
  Educational Question Generation (EQG) aims to synthesize customized exercise questions that enhance student learning. An effective EQG system should ideally personalize questions for each student by modeling the student’s knowledge state and generating questions that provide the greatest learning benefit. However, few existing EQG approaches are able to achieve such fine-grained personalization.
  In this paper, we explore how EQG can benefit from knowledge tracing (KT), which models students’ knowledge states based on historical performance and predicts future performance. We propose \algname, a personalized EQG framework that generates effective questions for individual students under the guidance of a KT model. Specifically, \algname seeks to maximize a student’s potential improvement in overall knowledge mastery by leveraging the KT model to select the most suitable knowledge concept for the student to practice. An LLM-based question generator is then trained to produce a question faithfully grounded in the selected concept. Experimental results on \textsc{XES3G5M} and \textsc{MOOCRadar} show that \algname consistently generates more effective questions than methods with limited or no personalization.
\end{abstract}

\renewcommand{\thefootnote}{}
\footnotetext{Code available at \url{https://github.com/UCSB-NLP-Chang/KT4EQG}}
\renewcommand{\thefootnote}{\arabic{footnote}}

\section{Introduction}

Educational question generation (EQG) is the task that aims to automatically generate high-quality exercises \citep{kurdi2020systematic, mulla2023automatic,dong2025literature}. Students vary widely in their knowledge mastery. A question that is pedagogically valuable for one student may be too trivial, too difficult, or focused on already-mastered material for another. Therefore, an ideal EQG system should personalize questions based on the individual's knowledge state and learning stage. To achieve such personalization, the system must model students' knowledge states to estimate the value of questions for individuals. However, prior approaches have focused on text quality, question format, and difficulty control \citep{wang2018qg,wang2022towards,scaria2024automated,bi2024difficulty}, typically lacking consideration of how different questions might benefit a specific learner \citep{utami2024contextualized,lim2025learning}. 

\begin{figure*}[h]
    \centering
    \includegraphics[width=0.93\linewidth]{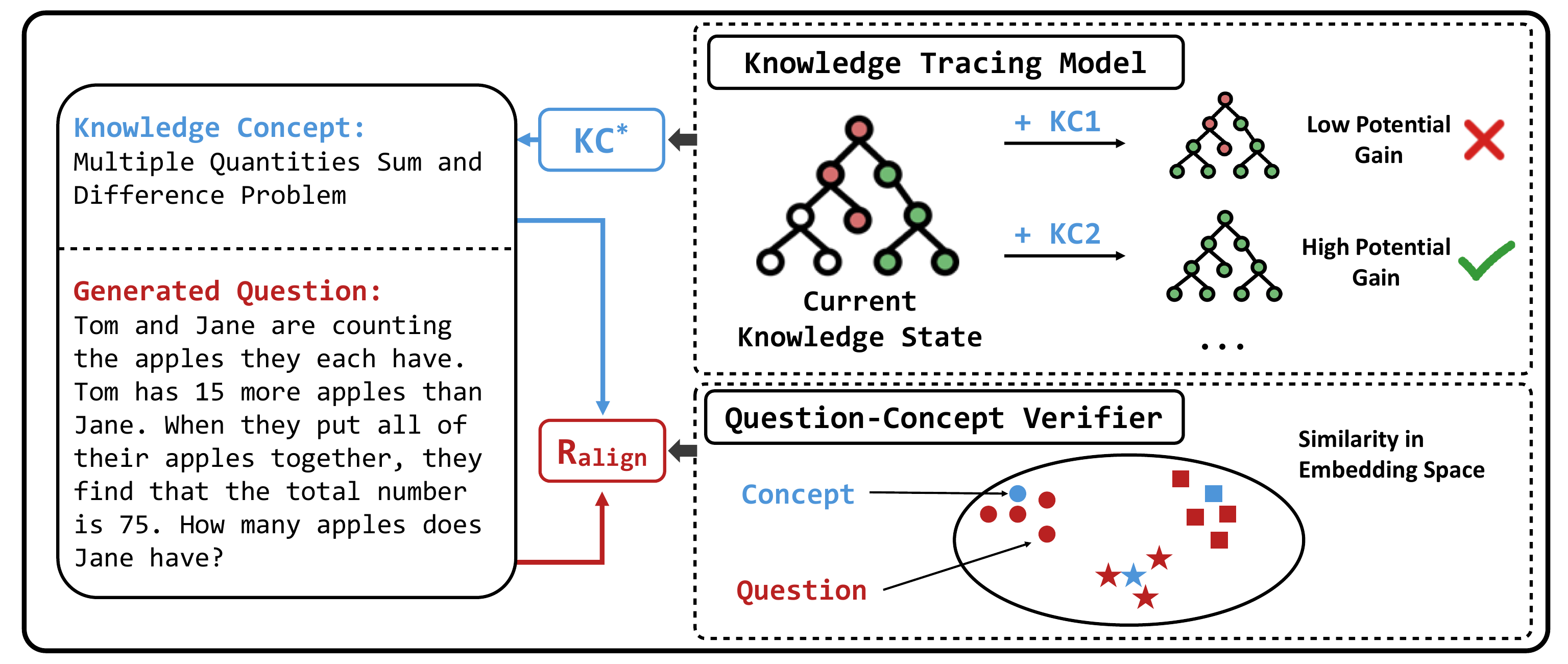}
    \caption{Overview of \algname. Left: An example of a KT-selected KC and a generated question. Right: The two components of the framework. The best KC is selected by maximizing the education value, which measures the potential improvement in the student's overall knowledge mastery after practicing a KC, as estimated by a tree-based knowledge tracer. $\mathcal{R}_\text{align}$ measures the question-concept alignment value by computing the similarity between the question and KC representations in the embedding space via a learned verifier.}
    \label{fig:main}
    \vspace{-4mm}
\end{figure*}

To model students' knowledge states for personalized EQG, knowledge tracing (KT) offers a natural approach. KT is an educational task that models a student's evolving knowledge state from their practice history and predicts future performance \citep{BKT,DKT,AKT}. A key characteristic of many KT methods is that they track knowledge mastery at the level of individual knowledge concepts (KCs), which represent specific skills and topics in a domain. This concept-level modeling provides concrete, assessable units that translate the abstract ``knowledge state" into measurable mastery levels, enabling precise identification of what each student knows and what they need to practice. Due to this capability, KT models have been increasingly used to support downstream educational decisions, such as question recommendation \citep{he2022exercise,pei2024enhanced,ExRec}, where knowledge tracers estimate students' learning benefits to support personalized selection of learning materials.

While prior work has leveraged KT models for personalizing exercise questions for each student, these methods focus on \textit{question recommendation} rather than \textit{question generation} -- they need to select questions from a given fixed exercise set.
In contrast, EQG focuses on a more challenging task: creating abundant, diverse questions for any target concept without being constrained by pre-existing questions. Yet the integration of KT with EQG to enable effective personalization remains barely explored.

In this work, we propose \algname, a personalized question generation framework guided by knowledge tracing. Instead of treating question generation as a purely text-based task, we explicitly incorporate student state modeling into the generation loop. Specifically, \algname first leverages a knowledge tracer to identify the most beneficial knowledge concept (KC) for the student to practice, and then uses a trained question generator to produce a question grounded in that concept. Toward this goal, we decompose question evaluation into two components: \ding{182} a \textbf{concept-level educational value} computed based on the knowledge tracer, which measures the potential improvement in the student's overall knowledge mastery after practicing, and \ding{183} a \textbf{question-concept alignment value} that measures whether the generated question actually tests the target KC. The concept with the highest educational value is directly selected by the KT model at inference time, while the question generator is trained via supervised fine-tuning, followed by reinforcement learning with a concept-alignment value as reward, ensuring the generated questions faithfully test the target KC.

Experimental results on \textsc{XES3G5M}~\citep{XES3G5M} and \textsc{MOOCRadar}~\citep{MOOCRadar} demonstrate that \algname generates personalized questions that are both pedagogically beneficial and concept-aligned, outperforming existing baselines. Ablation studies confirm the contribution of each component in the framework.

\section{Related Works}

\paragraph{Educational Question Generation and Data Synthesis.} 
Educational question generation has been studied for over a decade, with early EQG systems focusing on automated question generation leveraging statistical and semantic-based approaches~\citep{heilman2010good,yao2012semantics,chali2015towards}. Subsequent work shifted toward neural sequence-to-sequence models that learn question generation directly from data~\citep{du2017learning,zhou2017neural,kim2019improving}. More recently, pretrained transformer models and instruction-tuned large language models (LLMs) have further advanced EQG by enabling high-quality and controllable question generation through prompt engineering or domain-specific adaptation~\citep{wang2022towards,bulathwela2023scalable}. Other studies also focus on controlling specific attributes of generated questions, such as difficulty level or topic coverage~\citep{jiao2023automatic,li2025novel}. A related line of work uses question generation as a data-synthesis tool to improve language models' reasoning, particularly for mathematics~\citep{luo2023wizardmath,lu2024mathgenie,ding2024unleashing,zhao2025promptcot}. Although these methods are not designed for educational applications, they similarly aim to generate large-scale, diverse, and high-quality questions, sharing fundamental connections with EQG. Overall, these methods optimize question-level qualities and do not account for student-level personalization.

\paragraph{Knowledge Tracing.} Knowledge tracing models a student's evolving knowledge state from historical practice records and predicts future performance. A wide range of KT methods have been proposed, including probabilistic methods~\citep{BKT,IRT}, deep learning-based methods~\citep{DKT,AKT,GKT}, and LLM-based methods~~\citep{FT-LLM,LKT}. Several extensions further emphasize individualized knowledge modification by adapting learning dynamics~\citep{iBKT} or incorporating structured concept hierarchies~\citep{PSIKT,KT2}, enabling more personalized estimation of student knowledge state.

\paragraph{Personalized Question Recommendation and Generation.}
Previous work~\citep{jia2025eduagentqg,lim2025learning,utami2024contextualized} has explored generating personalized questions by incorporating student context, interests, or pedagogical feedback into the generation process. These methods emphasize contextual or stylistic personalization rather than fine-grained modeling of students' evolving knowledge states. To address fine-grained personalization, another line of work leverages KT to model students' mastery of knowledge and to support downstream educational decisions. KT-based methods have been increasingly studied for personalized question recommendation, where a knowledge model guides the selection of appropriate questions from a pre-given question set. Most of these methods apply KT to estimate students' concept mastery, and select exercises mainly based on predicted correctness, concept coverage, or concept-difficulty relationships~\citep{wu2020exercise,he2022exercise,pei2024enhanced}. More recent approaches enrich this process by incorporating semantic information of questions~\citep{wan2025exercise,ExRec}. In contrast, applying KT to open-ended question generation remains relatively limited, with existing efforts primarily using KT outputs for coarse difficulty adaptation or question filtering~\citep{srivastava2021question,yin2025leveraging}.

\section{Method}

\subsection{Task Formulation}
We begin by introducing our general notations. We use upper-cased letter, \emph{e.g.}, $\bm X$ or $X$, to denote random vectors/variables, and lower-cased letter, \emph{e.g.}, $\bm x$ or $x$, to denote deterministic vectors/scalars.

The personalized EQG problem is formulated as follows. During a student’s learning process, personalized EQG aims to generate questions that most effectively promote learning. Intuitively, an effective question should lead to measurable improvement in the student’s understanding within the target domain. In this work, we quantify this learning progress through a set of KCs, which represent fundamental units of knowledge to master. 

Formally, let $\mathcal{C}$ denote the set of KCs in a learning domain (e.g., a course or subject area), and let $\mathcal{K} = \bigcup_{c \in \mathcal{C}} \{K_c\}$ be a set of binary variables indicating whether the target student masters each KC. Let $\mathcal{H} = \bigcup_{t\in \mathcal{T}}\{(\bm X_t, A_t)\}$ be the set of historical questions for the target student, where $\bm X_t \in \mathcal{X}$ represents the content of the question $t$, and $A_t$ is a binary variable representing whether the target student answered question $t$ correctly. We then introduce the following \emph{mastery probabilities for the student},
\begin{equation}
    p(K_c = 1 | \mathcal{H}), \quad \forall c \in \mathcal{C},
\end{equation}
to track the student's knowledge mastery state after each practice question. As will be discussed in Section~\ref{sec:overview}, these probabilities can be estimated by a knowledge tracing algorithm.

Our goal is to generate the next practice KC and question for the student, $c_{t^*}$ and $\bm x_{t^*}$, respectively, such that the \emph{overall knowledge mastery} after practicing the question is maximized:
\begin{equation}
    \max_{\bm x_{t^*}} \sum_{c \in \mathcal{C}} p(K_c = 1 | \mathcal{H} \cup \{(\bm x_{t^*}, 1)\}) \equiv
    \max_{\bm x_{t^*}} \sum_{c \in \mathcal{C}} p(K_c = 1 | \mathcal{H}, \bm X_{t^*}=\bm x_{t^*}, A_{t^*}=1).
    \label{eq:utility}
\end{equation}

Here are two notes on the objective function in Eq.~\eqref{eq:utility}. First, we directly set the correctness of the new question, $A_{t^*}$, to 1, which means that we maximize the mastery probabilities assuming that the student can get the new question right. This is because in educational practice, exercises serve as learning opportunities rather than one-shot assessments: students can attempt problems, review solutions, and retry until mastery is achieved. 

Second, note that Eq.~\eqref{eq:utility} sums over \emph{all} the KCs, even though we are generating only one question at a time. This is because the knowledge concepts are correlated. Increasing the mastery of one KC is likely to impact the mastery of many other KCs. Hence, we want to generate a question that produces the most positive effects on all the KCs, rather than confining ourselves to the KCs to which the question is directly related.

Section~\ref{sec:overview} will elaborate how we optimize Eq.~\eqref{eq:utility} via a cascaded process.

\subsection{Overview of \algname}
\label{sec:overview}

To compute the objective function in Eq.~\eqref{eq:utility}, we adopt KT$^2$~\citep{KT2}, a KT algorithm based on a principled probabilistic model for tracking students KC mastery states. One core assumption in KT$^2$ is that each practice question $\bm X_t$ is associated with one KC, denoted as $\psi({\bm X_t})$ in this paper. We further assume that whether the student can answer $\bm X_t$ correctly (depicted by $A_t$) only depends on whether they master its underlying knowledge concept (depicted by $K_{\psi(\bm X_t)}$), \emph{i.e.,}
\begin{equation}
    p(A_t|\bm X_t, \{K_c\}_{c\in \mathcal{C}}) = p(A_t | K_{\psi(\bm X_t)}).
    \label{eq:suff_stat}
\end{equation}
In other words, $\psi(\bm X_t)$ becomes the \textit{sufficient statistics} of $\bm X_t$ for tracking the student's knowledge mastery level. Based on this assumption, Eq.~\eqref{eq:utility} can be further decomposed into
\begin{equation}
\begin{aligned}
    & \max_{\bm x_{t^*}} \sum_{c \in \mathcal{C}} p(K_c = 1 | \mathcal{H}, \bm X_{t^*} = \bm x_{t^*}, A_{t^*}=1) \\
    \Leftrightarrow &\max_{\bm x_{t^*}} \sum_{c \in \mathcal{C}} p(K_c = 1 | \mathcal{H}, \psi(\bm X_{t^*}) = \psi(\bm x_{t^*}), A_{t^*}=1) \\
    \Leftrightarrow & \max_{x_{t^*}, c_{t^*}} \underbrace{\Big[\sum_{c \in \mathcal{C}} p(K_c = 1 | \mathcal{H}, \psi(\bm X_{t^*}) = c_{t^*}, A_{t^*}=1)\Big]}_{\text{education value}}
    \cdot
    \underbrace{\mathbbm{1}[\psi(\bm x_{t^*}) = c_{t^*}] 
    \vphantom{\Big[\sum_{c \in \mathcal{C}} p(K_c = 1 | \mathcal{H}, \phi(\bm X_{t^*}) = c_{t^*}, A_{t^*}=1)\Big]}}_{\text{alignment}},
\end{aligned}
\label{eq:decompose}
\end{equation}
where $\mathbbm{1}[\cdot]$ is an indicator function which equals $1$ if the argument in its bracket is true, and $0$ otherwise. Eq.~\eqref{eq:decompose} breaks into two terms. The first term describes the educational benefit for the student if the KC of the new question is set to $c_{t^*}$, hence the name \emph{education value}; the second term describes whether the actual question $\bm x_{t^*}$ is truly aligned with the KC $c_{t^*}$, hence the name \emph{alignment}. This decomposition naturally leads to the following two-step optimization process:

\textbf{Step 1:} Find a KC for the new question, $c_{t^*}$, that maximizes the education value term.

\textbf{Step 2:} Generate a question $\bm x_{t^*}$ to align with the optimal $c_{t^*}$ (\emph{i.e.}, satisfying $\psi(\bm x_{t^*}) = c_{t^*}$).

Section~\ref{sec:edu} elaborates on Step 1, and Sections~\ref{sec:align} and \ref{sec:verifier} elaborate on Step 2.

\subsection{Maximizing Education Value}
\label{sec:edu}

We adopt an exhaustive search for the optimal KC. Specifically, for each candidate KC, we compute the education value as in Eq.~\eqref{eq:decompose}, and then choose the KC that leads to the largest education value. So the remaining question is how to compute the education value.

To this end, we adapt the inference process in KT$^2$. The detailed inference process in described in \citet{KT2} and Appendix~\ref{appx:kt2}. Here we only provide a brief introduction of the method. KT$^2$ builds a hidden Markov tree model for the joint distribution of two sets of variables, $\{K_c\}_{c \in \mathcal{C}}$ and $\{A_t\}_{t \in \mathcal{T}}$, where the former is a set of hidden variables, and the latter observed variables. Their joint distribution is decomposed as
\begin{equation}
    p\big(\{K_c\}_{c \in \mathcal{C}}, \{A_t\}_{t \in \mathcal{T}}\big) = p\big(\{K_c\}_{c \in \mathcal{C}}\big) \cdot p\big( \{A_t\}_{t \in \mathcal{T}} | \{K_c\}_{c \in \mathcal{C}}\big).
\end{equation}
The first term is called \emph{transition probabilities}. All the KCs are arranged in a tree format, where higher-level nodes represent broader KCs, and the child nodes correspond to the KCs that fall under the umbrella of their parent KC. Then the transition probabilities describe how the mastery of parent KC affects the mastery of the children KCs.

The second term is called \emph{emission probabilities}. It depicts how the correctness of question $t$ is affected by the mastery of its corresponding KC, namely $p(A_t | K_{\psi(\bm X_t)})$ as Eq.~\eqref{eq:suff_stat}. The detailed expressions for these probabilities can be found in Appendix~\ref{appx:kt2}.

With these probability assumptions, computing the education value boils down to computing the posterior distribution of the hidden variables, which can be derived with the Bayesian law. The computation can be efficiently performed using the standard upward-downward algorithm~\citep{HMT}. Further details are provided in Appendix~\ref{appx:kt2}.

\subsection{Maximizing Alignment}
\label{sec:align}

To generate a question $\bm x_t$ that aligns with a given KC, we train an LLM-based question generator, $f_\theta: \mathcal{C}\rightarrow \mathcal{X}$, via the following two phases.

\textbf{Phase 1: Supervised Fine-Tuning.} We first obtain a question dataset with KC labels, $\mathcal{D} = \{ \psi(\bm x_i), \bm x_i \}$, and fine-tune a text LLM to produce $\bm x_i$ given $\psi(\bm x_i)$ as the input by minimizing the cross-entropy loss.

\textbf{Phase 2: Reinforcement Learning with Alignment Reward.}
We then perform reinforcement learning using the following alignment reward $\mathcal{R}_\text{align}(\bm x, c)$, which measures the degree to which the generated question $\bm x$ aligns with KC $c$.
The reinforcement learning phase encourages the model to explore a broader range of question formulation, leveraging the LLM's generative capabilities to produce diverse questions while ensuring KC alignment.

The alignment reward is produced by a trained alignment verifier, see Section~\ref{sec:verifier} for details.

\subsection{Computing the Alignment Reward}
\label{sec:verifier}
We train a verifier $g_\phi:\mathcal{X}\times\mathcal{C}\to\mathbb{R}$, to determine the compatibility between question $x$ and concept $c$. As shown in Figure~\ref{fig:main} (right), the verifier maps a question $\bm x$ and its associated concept $c$ into a shared embedding space, with the objective of learning semantically consistent representations that bring aligned question–concept pairs closer together. As such, we optimize the verifier via an InfoNCE-style contrastive objective~\citep{oord2018representation}:
\begin{equation}
\mathcal{L}_{\text{align}}(\phi) = -\log \left(\frac{\exp(g_\phi(\bm x, c))} {\exp(g_\phi(\bm x,  c)) + \sum_{c^-}\exp(g_\phi(\bm x,  c^-))}\right),
\label{eq:infonce}
\end{equation}
where $c$ denotes the intended KC of the generated question $x$, and $c^-$ denotes concepts in the batch that are different from the intended concept $c$. 

To obtain numerically stable alignment scores while preserving relative preferences among candidates, we standardize the raw logits across all $c'$ for the same $x$. The alignment score is then computed as:
\begin{equation}
\mathcal{R}_{\text{align}}(\bm x,c) = \sigma\!\left(\frac{g_\phi(\bm x,c) - \mu_x}{\tau\,\sigma_x} \right),
\label{eq:verifier_impl_new}
\end{equation}
where $\mu_x$ and $\sigma_x$ denote the mean and standard deviation of $\{g_\phi(\bm x,c')\}_{c' \in \mathcal{C}}$, and $\tau$ is a temperature hyperparameter that controls score smoothness. This standardization ensures numerically stable scores while preserving relative ordering. 

\section{Experiment}

\subsection{Data Construction and Training Settings}

We conduct experiments on \textsc{XES3G5M}~\citep{XES3G5M} and \textsc{MOOCRadar}~\citep{MOOCRadar}. Both datasets require hierarchical KC annotations for KT$^2$. \textsc{XES3G5M} provides expert-annotated KC trees. While \textsc{MOOCRadar} provides only KCs and does not include such annotations, we adopt the LLM-constructed KC trees from \citet{KT2}, demonstrating that our framework can also be applied to datasets without human-annotated tree structures.

Our experiments focus on the three largest knowledge modules in \textsc{XES3G5M} as defined in the experimental setting of KT$^2$, selected based on the number of distinct KCs and exercises in the module. Compared to the original classroom-scale setting used in KT$^2$, we extend the number of students to 300 per module to support training and evaluation of the generator.

The KT$^2$ model is trained on each knowledge module to estimate students' latent knowledge states. For each student, the knowledge tracer outputs the posterior mastery probability $p(K_c = 1 \mid \mathcal{H})$ for each concept $c$, and predicts correctness probability $p(A_t = 1 \mid \mathcal{H}, \psi(X_t) = c)$ for practicing concept $c$, following the conditional independence assumption in Section~\ref{sec:overview}.
To model students at different learning progress, we construct multiple truncated histories for each student. Specifically, for each student's practice history $\mathcal{H}$, we record their first 10, 20, 30, 40, and 50 exercises as truncated histories $\mathcal{H}_t$, showing the practice record after $t$ exercises. Data statistics are shown in Appendix~\ref{appx:data}.
During training, we use these truncated histories to generate new questions and compute $\mathcal{R}_{\text{align}}$ as the reward for reinforcement learning.

\subsection{Experiment Setup}
\label{sec:setup}
We adopt \textsc{Qwen3-8B}~\citep{qwen3} as the base model for the question generator, trained via supervised fine-tuning followed by reinforcement learning. The alignment verifier is implemented using \textsc{all-MiniLM-L6-v2}~\citep{reimers2019sentence,wang2020minilm} as the backbone. As the verifier is frequently queried to compute rewards during reinforcement learning, we prioritize inference efficiency to enable scalable and stable training. 

\paragraph{Evaluation.} 
Our objective is to test how a student improves after learning from generated questions. Intuitively, high-quality generated questions should lead to better performance on subsequent questions, \emph{i.e.,} an ``exam''.
To do so, we first construct a fixed exam set by randomly sampling $n$ KCs with replacement from all candidates and retrieving one question per KC from the original dataset. For each truncated student history $\mathcal{H}_t$, we then simulate a $k$-round practice process. At each round, a question is generated for the student, and the student's knowledge state is updated by the KT model assuming successful practice. A key consideration is that the KC a student actually practices depends on the question's content, which may not match the method's intended KC. We therefore use the alignment verifier to identify the KC most aligned with the generated question, and use that KC for the state update. After $k$-round practice, we evaluate the student's updated knowledge state on the exam set to assess the method's effectiveness. See Appendix~\ref{appx:implement} for implementation details.

\paragraph{Metrics.} 
For the main results, we report the exam score and answerability. Exam score is defined as the average predicted correctness probability across all exam questions, computed by the KT model after the multi-round practice process. In our main experiments, we use KT$^2$ as the evaluation model; we further analyze the effect of using alternative KT models in Appendix~\ref{appx:KT}. The answerability measures whether the model generates a valid question with sufficient information to admit a well-defined solution. We evaluate answerability using \textsc{Qwen-3-4B}, which judges whether a generated question is solvable or suffers from missing conditions. For ablation studies, we additionally report question-concept alignment score ($\mathcal{R}_{\text{align}}$) and the negative log-likelihood (NLL) under the base \textsc{Qwen3-8B} model to assess alignment quality and whether the generated question remains fluent and natural.

{\renewcommand{\arraystretch}{1.1}
\begin{table*}[t]
  \small
  \centering
  \setlength{\tabcolsep}{3.5pt}
  \resizebox{\textwidth}{!}{
  \begin{tabular}{l|cc!{\vrule width 1pt}cc!{\vrule width 1pt}cc}
    \toprule
    \textbf{Model} 
    & \textbf{Exam Score} & \textbf{Ans.}
    & \textbf{Exam Score} & \textbf{Ans.}
    & \textbf{Exam Score} & \textbf{Ans.}\\
    \midrule

    \rowcolor[HTML]{D9E3F0} 
    \multicolumn{7}{c}{\textbf{\textsc{XES3G5M}}~\citep{XES3G5M}} \\
    \hline
    & \multicolumn{2}{c!{\vrule width 1pt}}{\textbf{Application Module}} 
    & \multicolumn{2}{c!{\vrule width 1pt}}{\textbf{Computation Module}}
    & \multicolumn{2}{c}{\textbf{Counting Module}}\\
    \hline

    \textsc{Initial} 
      & 0.6292 & -- 
      & 0.6045 & -- 
      & 0.6642 & -- \\

    \textsc{Random} 
      & 0.7855 & -- 
      & 0.7667 & -- 
      & 0.7894 & -- \\

    \textsc{Oracle} 
      & 0.8274 & -- 
      & 0.8221 & -- 
      & 0.8164 & -- \\
    \arrayrulecolor{gray!40}\hline\arrayrulecolor{black}

    \textsc{Qwen3-8B}~\citep{qwen3}
      & 0.6669 & 0.993 
      & 0.7036 & 0.954 
      & 0.7417 & 0.988 \\

    \textsc{Qwen3-8B (+Oracle)}
      & 0.7378 & 0.997 
      & 0.7381 & 0.948 
      & 0.7640 & 0.898 \\

    \textsc{PromptCoT-2.0}~\citep{zhao2025promptcot} 
      & 0.7624 & 0.891 
      & 0.7447 & 0.886 
      & \underline{0.7935} & 0.918 \\

    \textsc{PromptCoT-2.0 (+Oracle)} 
      & \underline{0.7828} & 0.874 
      & \underline{0.7591} & 0.840 
      & 0.7888 & 0.894 \\

    \textsc{ScaleQuest-Qwen2-Math-7B-QGen}~\citep{ding2024unleashing} 
      & 0.7614 & 0.978 
      & 0.7391 & 0.993 
      & 0.7808 & 0.978 \\

    \textsc{ScaleQuest-Qwen2-Math-7B-QGen (+Oracle)} 
      & 0.7643 & 0.971
      & 0.7452 & 0.984 
      & 0.7927 & 0.978 \\

    \rowcolor{gray!20}
    \textsc{KT4EQG} 
      & \textbf{0.8036} & 0.993 
      & \textbf{0.7811} & 0.969 
      & \textbf{0.8128} & 0.955 \\

    \midrule

    \rowcolor[HTML]{D9E3F0} 
    \multicolumn{7}{c}{\textbf{\textsc{MOOCRadar}}~\citep{MOOCRadar}} \\
    \hline
    & \multicolumn{2}{c!{\vrule width 1pt}}{\textbf{Wine Knowledge}} 
    & \multicolumn{2}{c!{\vrule width 1pt}}{\textbf{Circuit Design}}
    & \multicolumn{2}{c}{\textbf{Education Theory}}\\
    \hline

    \textsc{Initial} 
      & 0.6971 & -- 
      & 0.7210 & -- 
      & 0.6865 & -- \\

    \textsc{Random} 
      & 0.7783 & -- 
      & 0.8260 & -- 
      & 0.9173 & -- \\

    \textsc{Oracle} 
      & 0.9122 & -- 
      & 0.8345 & -- 
      & 0.9603 & -- \\

    \arrayrulecolor{gray!40}\hline\arrayrulecolor{black}

    \textsc{Qwen3-8B}~\citep{qwen3}
      & 0.8312 & 0.968 
      & 0.7809 & 0.997 
      & 0.8906 & 1.000 \\

    \textsc{Qwen3-8B (+Oracle)}
      & 0.8292 & 0.995 
      & 0.7449 & 1.000 
      & \underline{0.8909} & 1.000 \\

    \textsc{PromptCoT-2.0}~\citep{zhao2025promptcot} 
      & \underline{0.8454} & 0.921 
      & 0.7705 & 0.934 
      & 0.6546 & 0.962 \\

    \textsc{PromptCoT-2.0 (+Oracle)} 
      & 0.8421 & 0.938 
      & 0.8106 & 0.919 
      & 0.8630 & 0.969 \\

    \textsc{ScaleQuest-Qwen2-Math-7B-QGen}~\citep{ding2024unleashing} 
      & 0.7827 & 0.886 
      & 0.7597 & 0.987 
      & 0.8669 & 0.956 \\

    \textsc{ScaleQuest-Qwen2-Math-7B-QGen (+Oracle)} 
      & 0.8235 & 0.959 
      & \underline{0.8199} & 0.987 
      & 0.8253 & 0.992 \\

    \rowcolor{gray!20}
    \textsc{KT4EQG} 
      & \textbf{0.8789} & 0.928 
      & \textbf{0.8319} & 0.963 
      & \textbf{0.9181} & 0.979 \\

    \bottomrule
  \end{tabular}
  }
  \caption{Results on all knowledge modules. \textsc{Initial}, \textsc{Random}, and \textsc{Oracle} do not involve actual question generation and are reported for reference. These three baselines are excluded from the ranking. The best results are in \textbf{bold} and the second best are \underline{underlined}. \algname achieves the best performance across all modules on both datasets.}
  \label{tab:main-results}
  \vspace{-4mm}
\end{table*}
}

\paragraph{Baselines.}
We compare our method with the following baselines, including pre-trained LLMs and math question generation models:

$\bullet$ \textsc{Initial}: The simulated exam score before any practice, serving as a lower bound.

$\bullet$ \textsc{Random}: selects a KC uniformly at random at each practice round and updates the student state, serving as a naive baseline that operates independently of student's mastery.

$\bullet$ \textsc{Oracle}: Select the KC that maximizes the education value term (Eq.~\ref{eq:decompose}) at each practice round via exhaustive search over all candidate KCs and update the student state, serving as an upper bound that assumes perfect concept alignment of the practice questions.

$\bullet$  \textsc{Qwen3} (8B)~\citep{qwen3}: A widely-used pretrained instruction-tuned model with strong instruction-following capabilities. Evaluated to assess zero-shot performance.

$\bullet$  \textsc{ScaleQuest} (7B)~\citep{ding2024unleashing}: A scalable data synthesis method using lightweight 7B-scale question generators trained via two-stage fine-tuning to generate diverse mathematical questions from scratch.

$\bullet$  \textsc{PromptCoT-2.0} (30B)~\citep{zhao2025promptcot}: A concept-conditioned prompt synthesis framework that iteratively refines rationales via expectation-maximization to guide question construction, producing questions aligned with the specified concepts.

For all question-generating baselines, we report two variants: the default setting where the model selects its own KC from the candidate set and then generates an associated question, and a +\textsc{Oracle} setting where the KC selected by KT$^2$ is provided as a prefix, so that the model only needs to generate a question aligned with the given concept. This isolates the effect of question generation quality from concept selection.

\subsection{Main Result}

Table~\ref{tab:main-results} shows the performance of all methods across all knowledge modules in \textsc{XES3G5M} and \textsc{MOOCRadar}. \algname achieves the highest exam score across all modules, consistently outperforming all question-generation baselines. 

On \textsc{XES3G5M}, \algname surpasses the strongest baseline and approaches the \textsc{Oracle} upper bound, demonstrating that our framework effectively combines beneficial concept selection with well-aligned question generation. Notably, providing baselines with oracle KC selection (+\textsc{Oracle}) substantially improves their performance in most cases, confirming that concept selection is a key bottleneck for existing methods. However, even with oracle KCs, baselines still perform worse than \algname, indicating that our alignment-trained generator produces higher-quality questions for target KC. Results on \textsc{MOOCRadar} further validate the generalizability of \algname. Despite relying on LLM-constructed rather than human-annotated KC trees, \algname still achieves the best performance.    

Meanwhile, we observe that \algname consistently produces valid and answerable questions. Specifically, Table~\ref{tab:main-results} shows that the answerability of the generated questions remains comparable to that of the base model, \textsc{Qwen3-8B}, as well as other baseline models. This suggests that the proposed framework preserves question quality while optimizing for educational objectives.

To verify that our results are not specific to the choice of the evaluation model, we additionally evaluate all generation methods using BKT~\citep{BKT} and DKT~\citep{DKT} as alternative knowledge tracers. Results (Appendix~\ref{appx:KT}) show that \algname remains competitive, though the advantage varies across modules, which is expected as different knowledge tracers differ substantially in their modeling assumptions and capacity.

\subsection{Ablation Study}
\label{sec:ablation}
Table~\ref{tab:ablation-single-module} shows the progressive improvement of \algname through each training phase on the Application Module. To ensure a fair comparison, here we evaluate questions generated for the truncated student history without multi-round practice, so that all models operate on identical knowledge states. Starting from the backbone \textsc{Qwen3-8B} model, SFT Warmup fine-tunes on the training data, enabling the model to generate questions that resemble the dataset's question types and styles, thereby improving $\mathcal{R}_{\text{align}}$. Alignment RL then further improves concept alignment substantially, demonstrating the effectiveness of reinforcement learning with learned alignment reward. The NLL under the base model increases moderately after training, indicating a slight shift in generation distribution, but remains at a level that produces fluent and natural questions. Meanwhile, the answerability of the generated questions remains high throughout all training stages, indicating that our fine-tuning process does not degrade question quality.

\begin{table}[h]
\small
\centering
\setlength{\tabcolsep}{4pt}
\renewcommand{\arraystretch}{1.1}
\begin{tabular}{p{0.3\columnwidth}ccccc}
\toprule
\textbf{Setting} 
& \textbf{$\mathcal{R}_{\text{align}}$} 
& \textbf{NLL} 
& \textbf{Ans.} \\
\midrule
Backbone
  & 0.7845 & 1.4849 & 0.983 \\
SFT Warmup 
  & 0.8790 & 2.3226 & 0.990 \\
Alignment RL 
  & 0.9444 & 2.2662 & 0.997 \\
\bottomrule
\end{tabular}
\caption{Progressive performance. Each row shows cumulative results after the phase.}
\label{tab:ablation-single-module}
\vspace{-4mm}
\end{table}

\begin{figure*}[h]
  \centering
  \begin{minipage}[t]{0.49\linewidth}
    \centering
    \includegraphics[width=\linewidth,trim=0 0 0 0, clip]{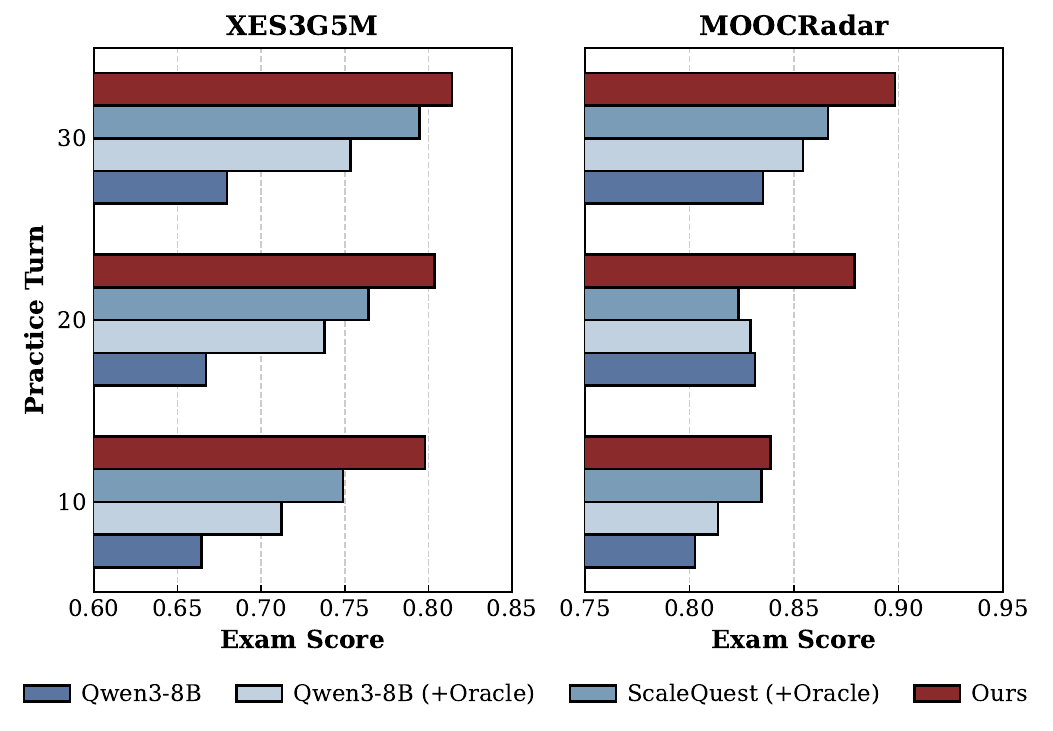}
  \end{minipage}
  \hfill
  \begin{minipage}[t]{0.49\linewidth}
    \centering
    \raisebox{1mm} {\includegraphics[width=\linewidth,trim=0 0 0 0, clip]{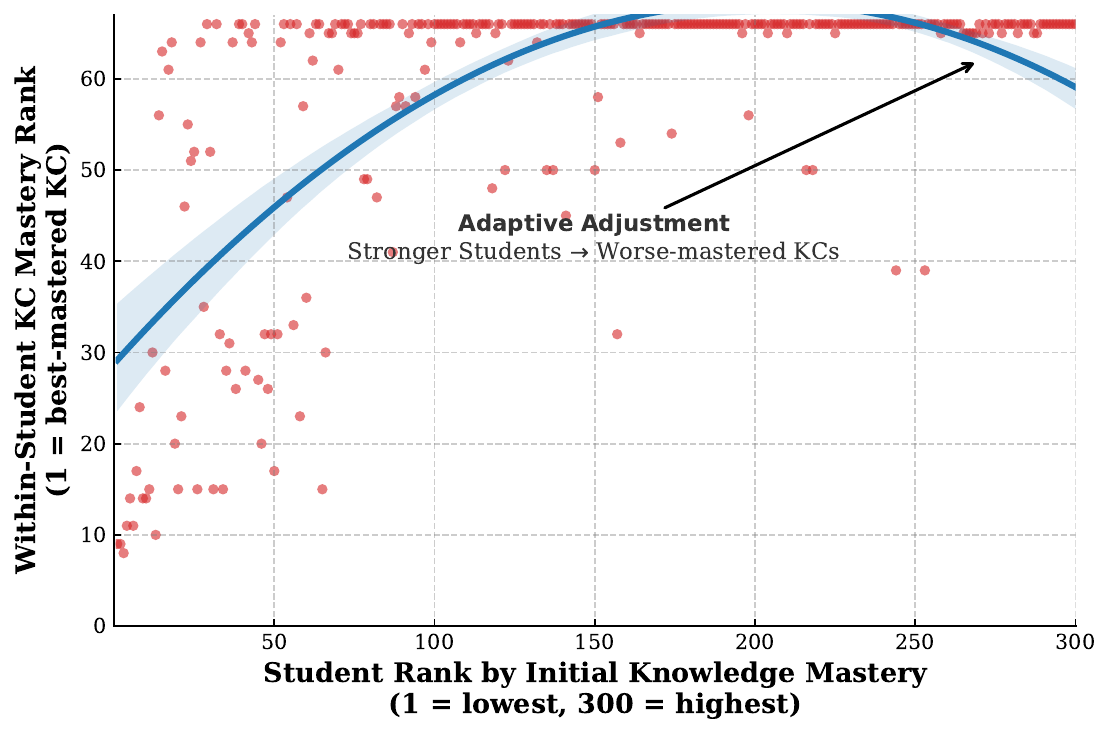}}
  \end{minipage}

  \vspace{-2mm}
  
  \begin{minipage}[t]{0.49\linewidth}
    \captionof{figure}{Effect of practice rounds on exam score. Results are shown for k=10,20,30. Methods with effective concept selection show greater gains with more practice.}
    \label{fig:multi-turn}
  \end{minipage}
  \hfill
  \begin{minipage}[t]{0.49\linewidth}
    \captionof{figure}{Adaptive Difficulty Adjustment. Students are ranked in ascending order of initial knowledge mastery. For each student, the selected KC is assigned a mastery rank.}
    \label{fig:adaptive_difficulty}
  \end{minipage}
  \vspace{-5mm}
\end{figure*}

We further study the effect of varying the number of practice rounds on exam scores. Figure~\ref{fig:multi-turn} reports results for $k=10,20,30$ on the representative module from both datasets. All methods improve with more practice rounds, consistent with the assumption that students practice until mastery at each round, but the rate of improvement differs across methods. Without oracle KC selection, \textsc{Qwen3-8B} shows only marginal gains as practice rounds increase, suggesting that poorly selected KCs provide limited cumulative learning benefit even with more practice. In contrast, methods with oracle KC exhibit substantially greater improvement. \algname achieves the highest exam score across all practice rounds, demonstrating consistent learning gains across varying practice times. 

\subsection{KC Selection Strategy Analysis}
To better understand the KC selection behavior driven by the objective (Eq.~\ref{eq:decompose}), we analyze the student mastery levels of selected KCs. We conduct this analysis on the Application Module, examining the KC selected for each truncated student history to isolate the selection behavior from subsequent simulated state updates. Figure~\ref{fig:adaptive_difficulty} reveals how \algname adapts KC selections to students' capabilities. We rank students according to their initial knowledge state in ascending order. Then, for each student, we rank all 66 candidate KCs by their mastery probability for that student, and plot the mastery rank of the KC selected by maximizing the education value. This mastery ranking indicates how well the student masters a KC compared to other candidates, with rank 1 corresponding to the KC with the highest mastery. The smoothed trend shows a clear adaptive pattern. 

For weaker students, the selection tends to favor KCs with higher mastery levels and avoid concepts that are too difficult. For students with higher mastery, the selection progressively shifts toward more challenging KCs. Overall, our selection naturally produces an adaptive policy that appropriately challenges students based on their current learning state.

\subsection{Question-Concept Alignment Case Study}

To examine whether alignment training produces questions that faithfully test the target KC, we compare questions generated by \algname and \textsc{Qwen3-8B (+Oracle)} on the Application Module. Since both methods receive the same oracle KC before question generation, differences in output quality reflect alignment capability rather than concept selection. For concept \textit{Multiple Quantities Chicken-Rabbit Problem}, \algname generates: "There are a total of 23 motorcycles and cars in the parking lot, with 54 wheels in total. How many cars are there?", which is a standard chicken-rabbit problem that requires reasoning over two entity types with different attributes. \textsc{Qwen3-8B (+Oracle)} generates: "If the total number of apples in a basket is 24 and the number of red apples is 10, how many green apples are there?", again reducing to simple subtraction. Additional examples comparing all baselines are provided in Appendix~\ref{appx:case_study}.

Besides, during the multi-round practice simulation, we observed that when the verifier identifies the KC that each generated question tests, the most frequent verified KC accounts for $5.2\%$ of \algname, compared to  $35.3\%$ for \textsc{Qwen3-8B (+Oracle)}, $8.7\%$ for \textsc{PromptCoT-2.0 (+Oracle)}, and $7.0\%$ for \textsc{ScaleQuest (+Oracle)}. The high concentration in the \textsc{+Oracle} variant of baselines indicates that without alignment training, generated questions may collapse to similar content regardless of the intended KC.

\section{Conclusion}
In this work, we address the challenge of personalized educational question generation by integrating the knowledge tracing method into the generation process. We propose \algname, a framework that generates pedagogically effective exercises tailored to learners' knowledge states. 
\algname decomposes the question evaluation into concept-level educational value and question-concept alignment value. We utilize the KT model to select the optimal concept at inference time, while training the question generator via supervised fine-tuning followed by reinforcement learning with a learned alignment reward to ensure accurate concept grounding.  

Experimental results show that \algname consistently generates questions that are both concept-aligned and beneficial for student learning, compared to existing baselines. Overall, this work demonstrates the potential of integrating knowledge tracing with question generation to support personalization in educational systems.

\section*{Acknowledgments}
UCSB acknowledges the support from the National Science Foundation (NSF) Grant IIS-2338252. Additionally, Xinyi Gao, Qiucheng Wu, Ying Xu, and Shiyu Chang acknowledge support from NSF Grant IIS-2302730.

\bibliography{colm2026_conference}
\bibliographystyle{colm2026_conference}

\appendix

\section{KT$^2$ Model Details}
\label{appx:kt2}

We briefly describe the KT$^2$ model~\citep{KT2} used in our framework. KT$^2$ builds a Hidden Markov Tree Model over the hierarchical KC structure. For each student, the hidden variables $\{K_c\}_{c \in \mathcal{C}}$ represent concept mastery, and the observed variables correspond to the student's correctness on individual exercises. The joint distribution decomposes into \textit{transition probabilities} and \textit{emission probabilities}.

\paragraph{Transition Probabilities.}
The mastery variables $\{K_c\}_{c \in \mathcal{C}}$ follow the topological structure of the KC tree. Each $K_c$ depends only on the mastery of its parent concept $\mathcal{P}(c)$:
\begin{equation}
    p\!\left(\{K_c\}_{c \in \mathcal{C}}\right) = \prod_{c \in \mathcal{C}} p(K_c \mid K_{\mathcal{P}(c)}).
\end{equation}
The transition probability is defined based on the assumption that mastering a parent KC entails mastering all its children (but not vice versa):
\begin{equation}
    p(K_c = 1 \mid K_{\mathcal{P}(c)} = m) = 
    \begin{cases}
        1 & \text{if } m = 1, \\
        \gamma_c & \text{otherwise},
    \end{cases}
\end{equation}
where $\gamma_c$ is a learnable parameter. For the root node, the condition on $K_{\mathcal{P}(c)}$ is removed and $\gamma_c$ serves as the root prior mastery probability.

\paragraph{Emission Probabilities.}
Each observed correctness variable $A_t$ depends only on the mastery of its associated KC $\psi(X_t)$:
\begin{equation}
    p(A_t = 1 \mid K_{\psi(X_t)} = m) = 
    \begin{cases}
        r_d & \text{if } m = 1, \\
        \varepsilon & \text{otherwise},
    \end{cases}
\end{equation}
where $r_d \in \{r_{\text{easy}}, r_{\text{med}}, r_{\text{hard}}\}$ is the correct probability given mastery, determined by the difficulty level $d$ of the exercise, and $\varepsilon$ represents the probability of answering correctly by guessing. The constraint $\varepsilon < r_{\text{hard}} < r_{\text{med}} < r_{\text{easy}}$ is enforced during parameter estimation. In our framework, since we select only the target KC without specifying difficulty, we fix $d = \text{med}$ for all KT-based evaluations, i.e., $r_d = r_{\text{med}}$.

\paragraph{Parameter Estimation.}
The model parameters $\theta = [\{\gamma_c\}_{c \in \mathcal{C}},\, r_{\text{easy}},\, r_{\text{med}},\, r_{\text{hard}},\, \varepsilon]$ are estimated via the Expectation-Maximization algorithm~\citep{EM}. The model also supports incremental updates during inference time, as new student responses are observed. See \citet{KT2} for details.

\paragraph{Inference.}
Given a student's practice history $\mathcal{H}$, we compute the posterior mastery probability $p(K_c = 1 \mid \mathcal{H})$ for each concept $c$. This can be decomposed as:
\begin{equation}
    p(A_t = 1 \mid \mathcal{H}, \psi(X_t) = c) = p(K_c = 0 \mid \mathcal{H})\cdot\varepsilon + p(K_c = 1 \mid \mathcal{H})\cdot r_d,
\end{equation}
where the posterior $p(K_c = 1 \mid \mathcal{H})$ is efficiently computed via the upward-downward algorithm on the KC tree~\citep{HMT}. This posterior mastery probability is what our framework uses to compute the education value term (Eq.~\ref{eq:decompose}) and to update student states during the multi-round evaluation.

\section{Data Statistics}
\label{appx:data}
\textsc{XES3G5M}~\citep{XES3G5M} contains 18,066 unique students, 7,652 questions, and 865 KCs, including both leaves and intermediate concepts. The dataset provides expert-annotated hierarchical KC structures, where questions are associated with leaf KCs only. \textsc{MOOCRadar}~\citep{MOOCRadar} contains 14,224 students, 2,513 questions, and 5,600 fine-grained KCs. This dataset does not provide KC tree structures, so we use the classroom modules from \citet{KT2} where the KC hierarchy is annotated by LLM. 

We list the statistics of the six knowledge modules used in our experiments in Table~\ref{tab:dataset-stats}, where candidate KCs refer to all leaf KCs in the module tree. For all experiments, we perform a student-level split of data, using an 8:2 ratio to construct the training and test sets. All truncated histories recorded from the same student are assigned to the same split to prevent information leakage.

{\renewcommand{\arraystretch}{1.2}
\begin{table}[h]
  \small
  \centering
  \setlength{\tabcolsep}{4pt}
  \begin{tabular}{lcccc}
    \toprule
    \textbf{Module} & \textbf{Students} & \textbf{Truncated Histories} & \textbf{Candidate KCs} \\
    \midrule
    Application States   & 300 & 1,500  & 66 \\
    Computation States   & 300 & 1,500  & 40 \\
    Counting States   & 300 & 1,500  & 28 \\
    Wine Knowledge   & 300 & 1,500  & 4 \\
    Circuit Design   & 260 & 1,300  & 8 \\
    Education Theory   & 130 & 650  & 3 \\
    \hline
  \end{tabular}
  \caption{Dataset statistics.}
  \label{tab:dataset-stats}
\end{table}
}

\section{Implementation Details}
\label{appx:implement}
\subsection{Alignment Verifier}
The alignment verifier is trained for 70 epochs with batch size 64, learning rate $1e-3$, and temperature $\tau$ set to 0.07, on a single NVIDIA GeForce RTX 3090 Ti GPU (24GB). For each $(c,x)$ pair, we utilize the KC tree annotation in the dataset to randomly sample a sibling concept of $c$ and treat it as a hard negative. These hard negatives are combined with in-batch negatives during training.

\subsection{Alignment Training}

\begin{table}[h]
\centering
\small
\begin{tabular}{lcc}
\toprule
\textbf{Hyperparameter} & \textbf{SFT} & \textbf{RL} \\
\midrule
Epochs & 10 & 3 \\
Effective batch size & 32 & 32 \\
Learning rate & $2e-5$ & -- \\
KL coefficient & -- & 0.3 \\
Entropy coefficient & -- & 0.03 \\
Rollouts & -- & 4 \\
Max sequence length & 2048 & -- \\
Max prompt length & -- & 4608 \\
\bottomrule
\end{tabular}
\caption{Training hyperparameters.}
\end{table}

All phases are trained on 8 NVIDIA H100 GPU (80GB). We use GRPO~\citep{GRPO} as the RL algorithm. Although our formulation focuses on concept selection only, the KT$^2$ requires a difficulty input to compute predictions. In all KT-based evaluations, we fix the difficulty level to \texttt{medium}. 

Across all training phases, we apply gradient masking to restrict loss computation to the question text portion of the model's output. This allows us to use a consistent prompting format (See Appendix~\ref{appx:prompt}) that produces complete $(c,x)$ pairs in a structured JSON output, while ensuring learning signals are applied only to the question generation component. 

\subsection{Evaluation}
We use the publicly released checkpoints of all baseline models and perform inference with temperature 0.8, top-p 0.8, and a maximum generation length of 512. For all models, including \algname, when the output follows the expected JSON format, we extract the question text field directly. If JSON parsing fails, we remove the KC field from the raw output to prevent the oracle KC name from biasing the verifier, and use the remaining text as the generated question.

\section{Knowledge Tracer Generalization}
\label{appx:KT}

In our multi-round evaluation pipeline, the practice simulation is performed using KT$^2$ for all methods, as KT$^2$ supports efficient inference-time incremental updates that are well-suited for sequential state tracking. BKT and DKT are used only at the final evaluation stage, where they predict the student's correctness probability on the fixed exam set based on the accumulated practice history.

\begin{table*}[h]
  \small
  \centering
  \setlength{\tabcolsep}{5pt}
  \resizebox{\textwidth}{!}{
  \begin{tabular}{l|cc!{\vrule width 1pt}cc!{\vrule width 1pt}cc}
    \toprule
    \textbf{Model} 
    & \textbf{BKT} & \textbf{DKT}
    & \textbf{BKT} & \textbf{DKT}
    & \textbf{BKT} & \textbf{DKT}\\
    \midrule

    \rowcolor[HTML]{D9E3F0} 
    \multicolumn{7}{c}{\textbf{\textsc{XES3G5M}}} \\
    \hline
    & \multicolumn{2}{c!{\vrule width 1pt}}{\textbf{Application Module}} 
    & \multicolumn{2}{c!{\vrule width 1pt}}{\textbf{Computation Module}}
    & \multicolumn{2}{c}{\textbf{Counting Module}}\\
    \hline

    \textsc{Qwen3-8B}
      & 0.5799 & 0.7191 
      & 0.6994 & 0.8249 
      & 0.6375 & 0.7171 \\

    \textsc{Qwen3-8B (+Oracle)}
      & 0.5839 & \textbf{0.8057} 
      & 0.7041 & 0.8070 
      & 0.6163 & 0.7342 \\

    \textsc{PromptCoT-2.0} 
      & 0.6025 & 0.7807 
      & 0.7049 & 0.8289 
      & 0.6477 & 0.7100 \\

    \textsc{PromptCoT-2.0 (+Oracle)} 
      & 0.6066 & 0.7713 
      & \underline{0.7107} & 0.7996 
      & 0.6462 & 0.7241 \\

    \textsc{ScaleQuest-Qwen2-Math-7B-QGen} 
      & 0.6085 & \underline{0.7885} 
      & \textbf{0.7139} & \textbf{0.8685} 
      & 0.6408 & 0.6827 \\

    \textsc{ScaleQuest-Qwen2-Math-7B-QGen (+Oracle)} 
      & 0.6028 & 0.7611 
      & 0.7080 & \underline{0.8316} 
      & 0.6459 & 0.7287 \\

    \rowcolor{gray!20}
    \textsc{KT4EQG} 
      & \textbf{0.6202} & 0.7616 
      & 0.7083 & 0.8077 
      & \textbf{0.6696} & \textbf{0.7753} \\

    \midrule

    \rowcolor[HTML]{D9E3F0} 
    \multicolumn{7}{c}{\textbf{\textsc{MOOCRadar}}} \\
    \hline
    & \multicolumn{2}{c!{\vrule width 1pt}}{\textbf{Wine Knowledge}} 
    & \multicolumn{2}{c!{\vrule width 1pt}}{\textbf{Circuit Design}}
    & \multicolumn{2}{c}{\textbf{Education Theory}}\\
    \hline

    \textsc{Qwen3-8B}
      & 0.8197 & 0.9188 
      & 0.6487 & 0.5987 
      & 0.8478 & \textbf{0.9380} \\

    \textsc{Qwen3-8B (+Oracle)}
      & 0.8348 & 0.9186 
      & 0.6147 & 0.6076 
      & 0.7876 & \underline{0.9357} \\

    \textsc{PromptCoT-2.0}
      & 0.8359 & 0.9260 
      & \textbf{0.6818} & 0.6059 
      & 0.8760 & 0.8867 \\

    \textsc{PromptCoT-2.0 (+Oracle)}
      & 0.8312 & 0.9255 
      & 0.6697 & \underline{0.6304} 
      & \textbf{0.8871} & 0.8630 \\

    \textsc{ScaleQuest-Qwen2-Math-7B-QGen}
      & 0.8311 & 0.9149 
      & \underline{0.6721} & 0.5913 
      & \underline{0.8761} & 0.9142 \\

    \textsc{ScaleQuest-Qwen2-Math-7B-QGen (+Oracle)}
      & 0.8389 & 0.9209 
      & 0.6565 & 0.5987 
      & 0.8397 & 0.9059 \\

    \rowcolor{gray!20}
    \textsc{KT4EQG} 
      & \textbf{0.8727} & \textbf{0.9521} 
      & 0.6479 & \textbf{0.6564} 
      & 0.7899 & 0.9304 \\

    \bottomrule
  \end{tabular}
  }
  \caption{Generalization of generated questions evaluated under alternative KT models (BKT and DKT). The best results are in \textbf{bold} and the second best are \underline{underlined}.}
  \label{tab:kt-generalization}
\end{table*}

As shown in Table~\ref{tab:kt-generalization}, \algname remains competitive under both alternative KT evaluators. While the advantage is less consistent than the KT$^2$-evaluated results, this is expected as BKT and DKT differ substantially from KT$^2$ in their modeling assumptions and capacity. For instance, BKT assumes independent binary knowledge states, and DKT uses a flat sequential model. These architectural differences lead to different assessments of student mastery, which can affect the relative ranking of methods. Importantly, the alignment training component of our framework is entirely independent of KT method, and can be generalize to any knowledge tracer selection.

\newpage
\section{Question-Concept Alignment Case Study}
\label{appx:case_study}

Table~\ref{tab:alignment_case} presents representative generated questions for three KCs from the Application Module. All baseline methods use the +\textsc{Oracle} setting, where the target KC is provided by the KT model; method names omit this suffix for brevity. \ding{51}~indicates the generated question tests the target KC's core reasoning structure; \ding{55}~indicates it does not; $\triangle$~indicates partial alignment.

\begin{table}[h]
\centering
\small
\begin{tabular}{p{0.12\textwidth} p{0.12\textwidth} p{0.62\textwidth} c}
\toprule
\textbf{KC} & \textbf{Method} & \textbf{Generated Question} & \\
\midrule
\multirow{4}{*}{\parbox{0.12\textwidth}{\raggedright Arithmetic Sequence Application}}
& KT4EQG & There is a sequence of numbers, the 5th number is 13, the 9th number is 21, and each number is 4 more than the previous one. What is the 15th number in this sequence? & \ding{51} \\
& Qwen3-8B & If the ratio of two numbers is 3:4 and their difference is 5, what is the remainder when the smaller number is divided by the larger number? & \ding{55} \\
& ScaleQuest & There are 300 people in a stadium. After an accident, the number of people in the stadium is reduced by 30\%. After that, another 30 people left the stadium. How many people are remaining in the stadium? & \ding{55} \\
& PromptCoT & A clock strikes once at 1 o'clock, twice at 2 o'clock, three times at 3 o'clock, and so on. How many times does it strike in 15 minutes? & $\triangle$ \\
\midrule
\multirow{4}{*}{\parbox{0.12\textwidth}{\raggedright Single Unitization Problem}}
& KT4EQG & If 5 monkeys eat 25 peaches in 5 minutes, how many minutes will it take for 10 monkeys to eat 80 peaches? & \ding{51} \\
& Qwen3-8B & There are 24 apples in total. If I give away 6 apples, how many apples will I have left? & \ding{55} \\
& ScaleQuest & How many digits does the number 200 have? & \ding{55} \\
& PromptCoT & A company has 10 employees. In how many ways can a team of 3 employees be formed? & \ding{55} \\
\midrule
\multirow{4}{*}{\parbox{0.12\textwidth}{\raggedright Trees at Both Ends Problem}}
& KT4EQG & Eddie planted 5 trees on one side of a road, with the distance between each two trees being 3 meters. How long is this section of the road? & \ding{51} \\
& Qwen3-8B & There are 10 trees planted in a row, and the distance between each tree is 2 meters. What is the total distance from the first tree to the last tree? & $\triangle$ \\
& ScaleQuest & What is the sum of all two-digit numbers whose digits add up to 9? & \ding{55} \\
& PromptCoT & There is a row of trees by the river, with a distance of 5 meters between every two trees. There are 10 trees in total. What is the distance from the first tree to the last tree? & \ding{51} \\
\bottomrule
\end{tabular}
\caption{Question-concept alignment case study.}
\label{tab:alignment_case}
\end{table}

\section{Use of Large Language Models}\label{appx:LLM}
In addition to the core modeling, we utilized LLMs in the following aspects: \ding{182} The KC tree for \textsc{MOOCRadar} used in our experiments was sourced from \citet{KT2}, where the hierarchy was constructed using LLM-based annotation. \ding{183} Answerability of generated questions was evaluated using \textsc{Qwen-3-4B} as a judge, as detailed in Section~\ref{sec:setup}. \ding{184} LLMs were used to polish the writing and correct grammatical issues in the manuscript.

\newpage
\section{LLM Prompt}
\label{appx:prompt}

\begin{tcolorbox}[left=1.2pt,right=1.2pt,top=1.2pt,bottom=1.2pt]
\small
You are a helpful assistant that generates English questions for third grade students to support learning.\\
\\
You are given a set of candidate knowledge concepts and the student's current mastery levels. Your task is:\\
1) Select ONE knowledge concept that would be most helpful for the student to practice next, given their current mastery.\\
2) Generate exactly ONE English question for a third grade student that directly targets the selected knowledge concept.\\
\\
You must output a single JSON object following the format:\\
\{\\
``knowledge\_concept": ``...",\\
``question\_text": ``..."\\
\}\\
\\
Rules:\\
- Output MUST be a valid JSON object with exactly these two fields.\\
- Field names must match exactly: ``knowledge\_concept", ``question\_text".\\
- ``knowledge\_concept" must be chosen from the provided list.\\
- ``question\_text" must be in English, contain no answer, and match the knowledge\_concept.\\
- Do not output anything outside the JSON object.\\
\\
Knowledge Concepts and Student's Mastery Level:\\
\\
\textit{[Insert Student KC Mastery]}\\
\\
Choose exactly one knowledge concept from above for this student to practice.\\
Respond with the JSON object described in the instructions.\\
\end{tcolorbox}

\end{document}